\documentclass[preprint,aps,pra,groupedaddress,showpacs]{revtex4}
\usepackage{graphicx}
\begin{document}
\title{Comments on  the two-photon  interferometry}
\author {Fujio Shimizu}
\affiliation {Institute for Laser Science, University of Electro-Communications,
1-5-1 Chofugaoka, Chofu, Tokyo 113-8585, Japan}

\begin{abstract}
In this article we try to describe the physics of a standard 
optical interferometer fed by ``quantum'' photons in terms of
primitive, nevertheless accurate formulation.
We derive explicit interferene patterns and show how they vary
depending on the input photon state.

\end{abstract}

\maketitle

\section{Introduction}

Recent advance of optical technology has enabled us to generate
photon-number fixed states (Fock state of photons) and to observe
unusual interferometric phenomena such as Hong-Ou-Mandel dip 
and
fractional-period interference pattern
\cite{Hong,Fonseca,Ou,Edamatsu,Rarity,Beugnon,Kim2,Takesue,Aboussouan,
Xue,Jin,Jin2,Kim,Cai,Fakonas,Chen,Martino,Lopes,Jin3,Nagata,Olindo,
Heuer,Qiu}. 
Typical description of those phenomena is based on monochromatic
wave vectors which are stationary in time and extends to infinite space.
The advantage of this description is that 
the function of a delay line is described
by the phase shift of individual monochromatic waves. However, a photon
can be expressed based on any orthonormal set of wave vectors which
are contained in the structure of the relevant electro-magnetic wave field. 
Actual experiments use pulsed photons which occupy a finite space 
and change their wave form in time. 
In this article we try to describe the physics of an idealized two-path interferometer which is fed with one or two photons using the basis  
most convenient for describing experimental results.

In the next section we use the wave function of one of two incoming photons
as a basis vector of the orthnormal set describing the single photon space.. 
This description is convenient when we need at most two orthogonal vectors to describe the state of photons, and provides clear view of what
is expected to observe.

In the third section we return to the traditional way of description, and
discuss on phenomena when spectral characteristics of the photons
are important.

\subsection{Formulation}

A photon is a vector of norm (length) one in a Hilbert space, and can be
expressed by a complex function $E(\vec{r},l)$
in a three dimensional space, which satisfies
\begin{equation}
\sum_l \int d\vec{r} E(\vec{r},l)E^*(\vec{r},l)=1,
\end{equation}
where $\vec{r}$ is the three dimensional coordinates, and $l$ is additional
discrete variables to define the function $E$.

The inner product of the vector is defined by
\begin{equation}
<E_i,E_J>=\sum_l \int d\vec{r} E_i(\vec{r},l)E_j^*(\vec{r},l).
\end{equation}
Any vector $E_l(\vec{r},j)$ in the set is expressed as a sum of
orthonormal vectors $F_l(\vec{r},j)$ with a unitary matrix $s_{l,m}$,
\begin{equation}
E_l(\vec{r},j)=\sum_m s_{l,m}F_m(\vec{r},j).
\end{equation}

$F_j(\vec{r},l)$ develops in time satisfying Maxwell equations (in vaccum),
, and may change its shape $F_i(\vec{r},l,t)$ with time $t$. 
However, the inner product is
preserved, (therefore, together with the norm)  at any time,
\begin{equation}
<F_i(\vec{r},l,t),F_j(\vec{r},l,t)>=\delta_{i,j}
\end{equation}
Therefore, $F_j(\vec{r},l,t)$, $i=1,2,.....$, are the set of orthonormal 
photons at any time. Since $F_i(\vec{r},l,t)$ is a solution of Maxwell equation, 
the dynamics 
of single photon is the same as that of the classical electro-magnetic wave.

The difference between quantum and classical phenomena
occurs when two or more 
photons are involved. A two-photon state is not a vector of norm $\sqrt{2}$ in
the one-photon Hilbert space, but is a vector of norm one in the product space
of the two one-photon Hilbert space. In addition a vector in the 
two-photon space are restricted
to those which is symmetric with the exchange of photons 
in the two one-photon spaces,
\[
\alpha \biggl\{ E_i(\vec{r_1},l_1,t),E_j(\vec{r_2},l_2,t)
+E_i(\vec{r_2},l_2,t),E_j(\vec{r_1},l_1,t)\biggr\},
\]
for any $E_i$ and $E_j$, and $\alpha$ is the normalization factor.

The dynemics of a multi-photon interferometer is more conveniently
written by using photon-creation 
and annihilation operators. For elements of a set of orthonormal vectors
a creation operator $f_i^{\dagger}$ and an annihilation operator $f_j$  satisfy commutation relations
\begin{equation}
f_if_j^{\dagger}-f_j^{\dagger}f_i=\delta_{i,j},
\label{commutation}
\end{equation}
and $0$ for all other combinations. The photon vector is expressed 
as $f^{\dagger}|0>$ with the vacuum state $|0>$. Here, the suffix $i$ of
$f_i^{\dagger}$ specifies that the photon $f_i^{\dagger}|0>$ is the photon
$F_i(\vec{r},l,t)$ in function representation.

Any two-photon vector $|2>$ is expressed by a set of orthonormal
vectors as
\[
|2>=\biggl\{\sum_{i< j} s_{i,j}f_i^{\dagger} f_j^{\dagger}
+\sum_i s_{i,i}\frac{(f_i^{\dagger})^2}{\sqrt{2}}\biggr\}|0>,
\]
where \{$s_{i,j}$\}  is an unitary matrix.

Similarly we can expresse an n-photon vector $|n>$ by a sum of
orthonormal vectors in the $n$-photon space.
\begin{equation}
\frac{(f_{i_1}^{\dagger})^{n_1}(f_{i_2}^{\dagger})^{n_2}......
(f_{i_n}^{\dagger})^{n_n}}{\sqrt{n_1!n_2!......n_n!}}|0>
\label{normalizedn}
\end{equation}
with all combinations $(n_1,n_2,......,n_n)$ satisfying $n_1+n_2+......+n_n=n$.

\subsection{Approximation for two-path interferometer}

In an interferometer, mirrors, beam splitters, delay lines, 
phase plate, polarization plates and optical fibers change and restrict
the propagation of photons.  
We consider an idealized two-path interferometer in which a photon is 
a scalar wave traveling one dimensionally along two channels. 
The beam splitters mix waves of two channels. All optical elements
interact with a single photon instantaneously and locally. 
Furthermore, we restrict that all optical components including 
the transmission lines are dispersion free. 
In real world the above conditions are satisfied only by a limited range of
vectors of the system. If the photon propagate in free space, 
it expands gradually its cross-sectional size. Wavelength dispersion of
a beam splitter is not avoidable when the spectral range of the photon is
very large.  
However, we know that the above restrictions are satisfied in many
interferometric experiments, or at least we try to satisfy them in operation.
We restrict
the wave of a photon to,
\begin{equation}
E(\xi)=E_0(\xi)\mbox{e}^{ik_0\xi},
\label{quasimonochromatic}
\end{equation} 
where $\xi=x-ct$, and $E_0$ is a slowly varying scalar envelope function.

\begin{figure}[htbp]
\begin{center}
\includegraphics[width=12cm]{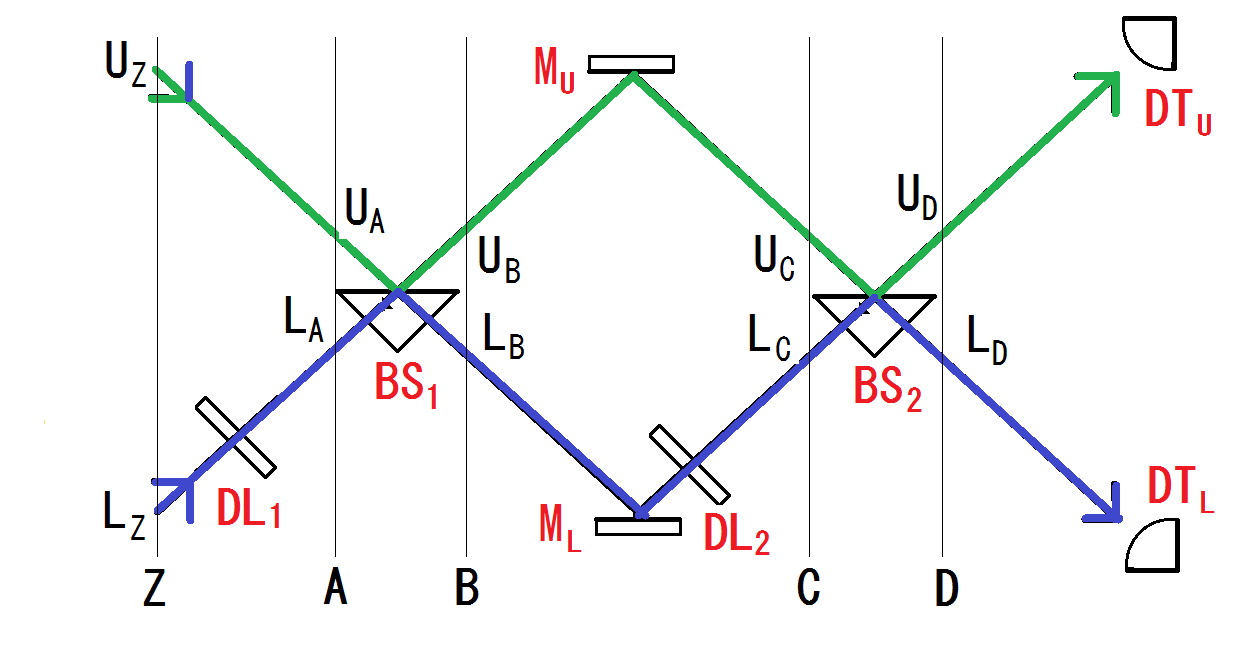}
\caption{The interferometer is composed of two one-dimensional 
channels U (green) and L (blue).
Two channels mix at two beam splitter BS$_1$ and BS$_2$.
In addition, in the lower channel L optical pulse go through one 
or two delay lines DL$_1$ and DL$_2$. Photons 
travelling the upper channel (U) is detected by DT$_U$, and the lower 
channel (L) by DT$_L$.}
\label{machzhender}
\end{center}
\end{figure}

Figure~\ref{machzhender} shows the system we discuss. 
Photons enter through the position Z or sometimes through A.
Amplitudes of the upper (U) and
the lower (L) channels mixes at two beam splitters. The state of the lower 
channel is modified by one or two delay lines.
The state of the photons is detected by detectors which are sensitive to
photons in either one of the two channels.
The relative spatial coordinate $\xi$ of the upper and lower channels
is defined to take the same value at the position Z (or sometimes 
at the beam splitter BS$_1$) and at the same time $t$.

The task is to express the state of photons at the input position Z or A by
orthonormal vectors which have amplitude only on the channel U or L at
the position of detectors, D or B..

\subsection{Detectors}

We place detectors which are sensitive only to the photon of one
of two channels. However, this does not mean that they are detecting
the magnitude of specific vector described in the article. 
A typical photo-electric 
detector consists of many photo-sensitive elements much smaller than the
transverse dimension of the optical wave. 
Intrinsic time resolution of the detector
may not match to the duration time of the photon. 
In this article otherwise stated we assume 
that the detector responds instantaneously.
Since we deal the photon which does not have transverse structure,
the instantaneous detector can detect all characteristics of 
arriving photons at the detector position.

Frequently the characteristics of the detector is modified by auxiliary components. They may produce phenomena which have little relation 
with the dynamics of the investigating interferometer. 
A typical example is a narrow-band filter. What the narrow-band detector 
does is that it divides an optical pulse into many paths, give different 
traveling length for each path, and then sum divided pulse components
together. It is a multi-path interferometer and produces interference
which does not exist in the investigating interferometer.

\subsection{Nomenclature}

We describe the creation (or annihilation)
operator of a photon 
whose amplitude is non-zero in a specific channel by $f_{i,X_W}^{\dagger}$. 
The first letter $i$ of the suffix
shows the wave form of the photon $E_i(\xi)$ which is the function of
the relative position of the wave along the channel $\xi=x-ct$.
The second letter is to show the channel this photon exists. It is U or L ,
meaning the upper or lower channel of the interferometer, respectively. 
The third letter W specifies the position in 
the interferometer. 
It runs Z (input to the system), A (input of the first beam splitter), 
B (exit of the first beam splitter), C (input of the second beam splitter), and 
D (the exit of the second beam splitter). 

To avoid confusion we use the word ''mode'' to specify the shape of the wave
function of the photon which enters the interferometer. Single mode does not 
mean that all photons propagating in the interferometer are an identical vector
of the Hilbert space of the interferometer. 
The vector is uniquely specified only by
the full specification of the creation (or annihilation) operator 
$f_{i,X_Y}^{\dagger}$ (or $f_{i,X_Y}$).    

\subsection{Beam splitter}{\label{subsecsplitter}}

\begin{figure}[htbp]
\begin{center}
\includegraphics[width=14cm]{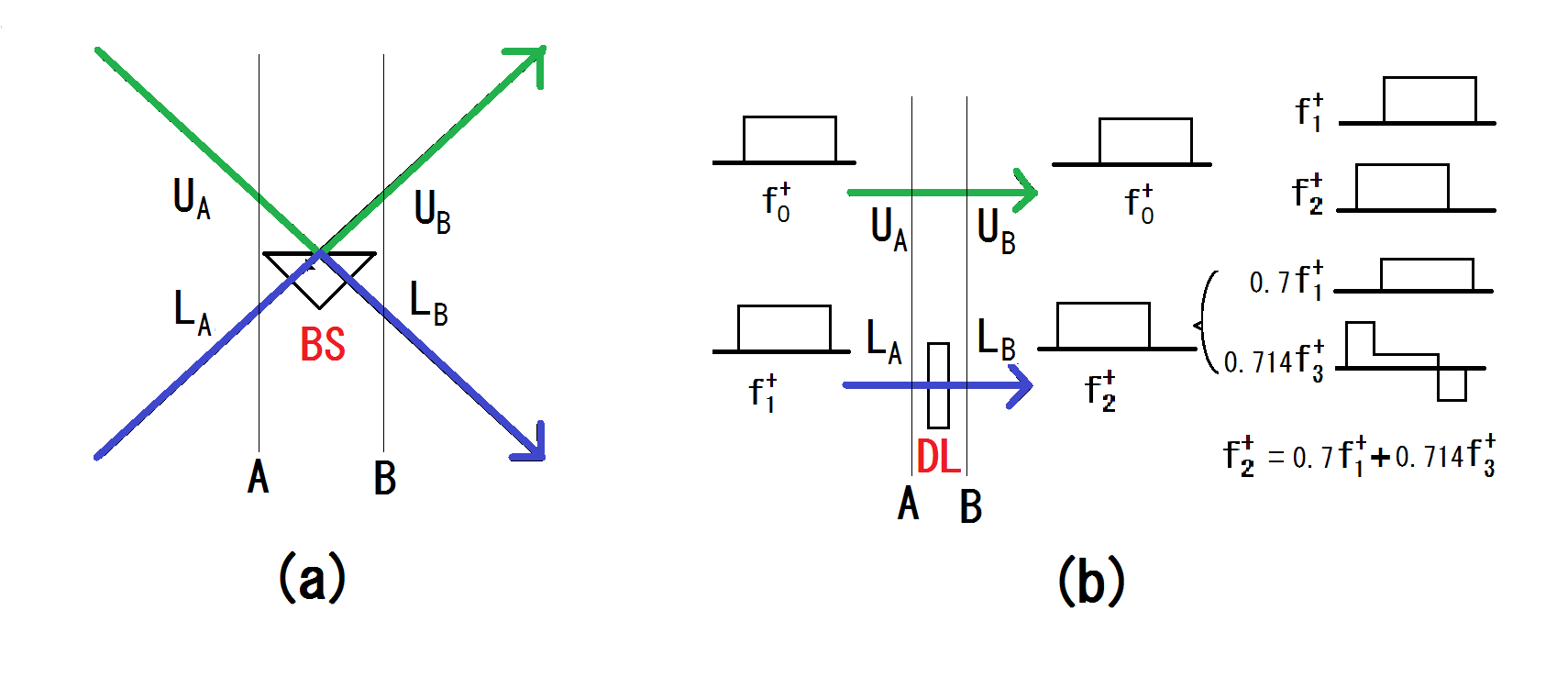}
\caption{Optical pulse propagation through (a): a 50\% beam splitter, and
(b): a delay line.}
\label{splitter}
\end{center}
\end{figure}

There are two components indispensable in an interferometer: two 
beam splitters and an variable delay line.
The function of a beam splitter is to mix amplitudes of wave functions in 
two channels.
When the beam splitter is lossless and divide amplitude equally,
the relation between the operator before the beam splitter (at the position A)
and after the beam splitter (at the position B) is
\begin{eqnarray}
\nonumber
f_{1,U_A}^{\dagger}|0>=
\frac{1}{\sqrt{2}}(f_{1,U_B}^{\dagger}+f_{1,L_B}^{\dagger})|0>
\\
f_{2,L_A}^{\dagger}|0>=
\frac{1}{\sqrt{2}}(f_{2,U_B}^{\dagger}-f_{2,L_B}^{\dagger})|0>.
\end{eqnarray}
Note, when $E_1(\xi)=\exp(i\phi)E_2(\xi)$, where $\phi$ is 
an arbitrary phase factor, two orthogonal vectors,
$f_{1,U}^{\dagger}|0>$ and $f_{1,L}^{\dagger}|0>$, are sufficient to describe
the state at the position B. However,
when $E_1(\xi)\neq\exp(i\phi) E_2(\xi)$,  
we need minimum of four mutually 
independent states,  $f_{1,U}^{\dagger}|0>$, $f_{2,U}^{\dagger}|0>$,
$f_{1,L}^{\dagger}|0>$, and $f_{2,L}^{\dagger}|0>$.

\subsection{Delay line}{\label{secshifter}}

The function of a delay line is the shift of the position $\xi$ of the
wave,
\begin{equation}
E_1(\xi)\rightarrow E_1(\xi+\Delta\xi)\equiv E_2(\xi).
\end{equation}
This inevitably change the vector $f_1^{\dagger}|0>$ to
a different vector $f_2^{\dagger}|0>$ which is not parallel to $f_1^{\dagger}|0>$. (See Fig.~\ref{splitter}(b).)
We decompose $f_2^{\dagger}|0>$
into the component parallel to the vector in the opposite channel,
$f_1^{\dagger}|0>$, and the component $f_3^{\dagger}|0>$ which is
perpendicular to $f_1^{\dagger}|0>$. 
Then, the vector in the upper and lower channels are
\begin{eqnarray}
\nonumber
f_{1,U_A}^{\dagger}|0>=f_{1,U_B}^{\dagger})|0>
\\
f_{1,L_A}^{\dagger}|0>= f_{2,L_B}^{\dagger}|0>=
\alpha_1 f_{1,L_B}^{\dagger}+\alpha_3 f_{3,L_B}^{\dagger})|0>,
\end{eqnarray}
where
\begin{equation}
\alpha_1=\int E_2(\xi)
E_1^*(\xi) d\xi
\label{delay}
\end{equation}
and $\alpha_3=\exp(i\phi)\sqrt{1-|\alpha_1|^2}$. The phase $\phi$ is arbitrary,
and $E_3(\xi)$ is determined by the equation 
\begin{equation}
E_2(\xi)=\alpha_1E_1(\xi)+\alpha_3E_3(\xi)
\end{equation}.

When $\Delta\xi$ is very small compared to the coherence length of
the photon pulse,  
the absolute value of the correlation is approximately one,
$|\alpha_1|\approx 1$. Then, the outgoing vector remains parallel 
to the incoming vector.
\begin{equation}
f_{1,L_B}^{\dagger}|0>=\mbox{e}^{ik_0\Delta\xi}f_{1,L_A}^{\dagger}|0>.
\label{phaseshift}
\end{equation}
Another case is the monochromatic wave
$E=\beta \exp(ik_0\xi)$, where $\beta$ is a constant.
In this case Eq.~(\ref{phaseshift}) is valid regardless of the length 
of the delay line.

\section{Mach-Zhender interferometer}\label{second}

When there is only one component in the interferometer,
which change the function 
shape of the photon, it is convenient to use one of the input photon
as one of the vector of the orthonormal set. 
Then, the interference is characterized 
with only one correlation function.

\subsection{single photon to channel U}

Let us first consider that a single photon enters the interferometer, which
should produce the interference pattern 
of a classical Mach-Zhender interferometer. 
Suppose, a photon $f_1^{\dagger}$ enter from the position  U$_A$. The state is
\begin{eqnarray}
\nonumber
f_{1,U_A}^{\dagger}=\frac{1}{\sqrt{2}}(f_{1,U_B}^{\dagger}+f_{1,L_B}^{\dagger})
=\frac{1}{\sqrt{2}}(f_{1,U_C}^{\dagger}+f_{2,L_C}^{\dagger})
=\frac{1}{\sqrt{2}}\biggl\{f_{1,U_C}^{\dagger}+(\alpha_1f_{1,L_C}^{\dagger}
+\alpha_3f_{3,L_C}^{\dagger})\biggr\}
\\
\nonumber
=\frac{1}{2}\biggl\{(f_{1,U_D}^{\dagger}+f_{1,L_D}^{\dagger})
+\alpha_1(f_{1,U_D}^{\dagger}-f_{1,L_D}^{\dagger})
+\alpha_3(f_{3,U_D}^{\dagger}-f_{3,L_D}^{\dagger})\biggr\}
\\
=\frac{1}{2}\biggl\{(1+\alpha_1)f_{1,U_D}^{\dagger}
+\alpha_3f_{3,U_D}^{\dagger}
+(1-\alpha_1)f_{1,L_D}^{\dagger}
-\alpha_3f_{3,L_D}^{\dagger}\biggr\}
\label{singleph}
\end{eqnarray}

In the above expression we omitted the vacuum vector $|0>$ at the end
of the expression. Because this does not cause confusion, we omit $|0>$
in the following sections.

Summing the intensity of all terms with U$_D$ or L$_D$, 
the probability of finding a photon in the channel U or L are 
\begin{eqnarray}
\nonumber
P_U=\frac{1}{4}(|1+\alpha_1|^2+|\alpha_3|^2)
=\frac{1}{2}+\frac{\alpha_1+\alpha_1^*}{4}
\\
P_L=\frac{1}{4}(|1-\alpha_1|^2+|\alpha_3|^2)
=\frac{1}{2}-\frac{\alpha_1+\alpha_1^*}{4},
\label{onephoton}
\end{eqnarray}
respectively.

The interference pattern is determined solely by the correlation $\alpha_1$
of the photon state before and after the delay line.
When the photon enters from the lower channel $L_A$, 
$P_U$ and $P_L$ must be exchanged.

\subsection{Two arbitrary photons through BS$_1$}
\label{mandelexperiment}

Next, let us discuss on the Hong-Ou-Mandel's experiment\cite{Hong} 
with the first 50\% 
beam splitter (BS$_1$ in Fig.~\ref{machzhender}).
The photon $f_{1,U_A}^{\dagger}$ enters from the position U$_A$, 
and the second photon $f_{2,L_A}^{\dagger}$ enters from 
the position L$_A$.
We decompose $f_{2,L_A}^{\dagger}$ into the  parallel $f_{1,L_A}^{\dagger}$ 
and perpendicular components $f_{3,L_A}^{\dagger}$ of 
$f_{1,L_A}^{\dagger}$ as
\begin{equation}
f_{2,L_A}^{\dagger}=\alpha_1f_{1,L_A}^{\dagger}
+\alpha_3f_{3,L_A}^{\dagger}
\end{equation}
(Note: $E_1(\xi)$ and $E_2(\xi)$ are not necessarily orthogonal, but
$f_{1,U_A}^{\dagger}|0>$ and $f_{2,L_A}^{\dagger}|0>$ are always
orthogonal because they travel in the different channel.)

Then, we try to write the two-photon state using the photons which have
amplitude only in the channel U$_B$ or the channel L$_B$.
The result is
\begin{eqnarray}
\nonumber
f_{1,U_A}^{\dagger}f_{2,L_A}^{\dagger}=f_{1,U_A}^{\dagger}
(\alpha_1f_{1,L_A}^{\dagger}+\alpha_3f_{3,L_A}^{\dagger})
\\
\nonumber
=\frac{1}{2}(f_{1,U_B}^{\dagger}+f_{1,L_B}^{\dagger})
\biggl\{\alpha_1(f_{1,U_B}^{\dagger}-f_{1,L_B}^{\dagger})
+\alpha_3(f_{3,U_B}^{\dagger}-f_{3,L_B}^{\dagger})\biggr\}
\\
\nonumber
=\frac{1}{2}\biggl\{\sqrt{2}\alpha_1\frac{(f_{1,U_B}^{\dagger})^2}{\sqrt{2}}
+\alpha_3f_{1,U_B}^{\dagger}f_{3,U_B}^{\dagger}
-\sqrt{2}\alpha_1\frac{(f_{1,L_B}^{\dagger})^2}{\sqrt{2}}
-\alpha_3f_{1,L_B}^{\dagger}f_{3,L_B}^{\dagger}
\\
-\alpha_3f_{1,U_B}^{\dagger}f_{3,L_B}^{\dagger}
+\alpha_3f_{3,U_B}^{\dagger}f_{1,L_B}^{\dagger}\biggr\}.
\label{mandelsplitter}
\end{eqnarray}
All terms in the last two lines of Eq.~(\ref{mandelsplitter}) are 
mutually orthogonal.
Using the above equation we obtain the probability having two atoms 
in U$_B$ or L$_B$,
$P_{UU}$ or $P_{LL}$, and the probability having a photon in both channels, 
$P_{UL}$.
\begin{eqnarray}
\nonumber
P_{UU}=P_{LL}=\frac{1}{2}|\alpha_1|^2+\frac{1}{4}|\alpha_3|^2
\\
P_{UL}=\frac{1}{2}|\alpha_3|^2
\end{eqnarray}

The coincidence probability $P_{UL}$ vanishes only when $E_1(\xi)$ is
equal to $E_2(\xi)$ excluding the global phase factor. In this case 
$|\alpha_1|=1$ and $\alpha_3=0$, and the state at the position B is
\begin{equation}
f_{1,U_A}^{\dagger}f_{1,L_A}^{\dagger}
=\frac{1}{2}\biggl\{(f_{1,U_B}^{\dagger})^2
-(f_{1,L_B}^{\dagger})^2\biggr\}
\label{mandeleq}
\end{equation}

\subsection{Two identical photons, one to each channel}

Let us now consider the case when two identical photons are sent to
the interferometer. The state at B is equal to Eq.~(\ref{mandeleq}).
Then,
\begin{eqnarray}
\nonumber
\frac{1}{2}\biggl\{(f_{1,U_B}^{\dagger})^2
-(f_{1,L_B}^{\dagger})^2\biggr\}
=\frac{1}{2}\biggl\{(f_{1,U_C}^{\dagger})^2
-(f_{2,L_C}^{\dagger})^2\biggr\}
\\
\nonumber
=\frac{1}{4}\biggl\{(f_{1,U_D}^{\dagger}+f_{1,L_D}^{\dagger})^2
-(\alpha_1f_{1,U_D}^{\dagger}+\alpha_3f_{3,U_D}^{\dagger}
-\alpha_1f_{1,L_D}^{\dagger}-\alpha_3f_{3,L_D}^{\dagger})^2\biggr\}
\\
\nonumber
=\frac{1}{4}\biggr\{\sqrt{2}(1-\alpha_1^2)\frac{(f_{1,U_D}^{\dagger})^2}
{\sqrt{2}}-2\alpha_1\alpha_3f_{1,U_D}^{\dagger}f_{3,U_D}^{\dagger}
-\sqrt{2}\alpha_3^2\frac{(f_{3,U_D}^{\dagger})^2}{\sqrt{2}}
\\
\nonumber
+\sqrt{2}(1-\alpha_1^2)\frac{(f_{1,L_D}^{\dagger})^2}{\sqrt{2}}
-2\alpha_1\alpha_3f_{1,L_D}^{\dagger}f_{3,L_D}^{\dagger}
-\sqrt{2}\alpha_3^2\frac{(f_{3,L_D}^{\dagger})^2}{\sqrt{2}}
\\
+2(1+\alpha_1^2)f_{1,U_D}^{\dagger}f_{1,L_D}^{\dagger}
+2\alpha_3^2f_{3,U_D}^{\dagger}f_{3,L_D}^{\dagger}
+2\alpha_1\alpha_3f_{1,U_D}^{\dagger}f_{3,L_D}^{\dagger}
+2\alpha_1\alpha_3f_{3,U_D}^{\dagger}f_{1,L_D}^{\dagger}\biggr\}.
\end{eqnarray}

The probabilities $P_{UU}$, $P_{LL}$, and $P_{UL}$ are
\begin{eqnarray}
\nonumber
P_{UU}=P_{LL}=
\frac{1}{8}|1-\alpha_1^2|^2+\frac{1}{8}|\alpha_3|^4
+\frac{1}{4}|\alpha_1\alpha_3|^2
=\frac{1}{4}-\frac{\alpha_1^2+(\alpha_1^*)^2}{8}
\\
P_{UL}=\frac{1}{4}|1+\alpha_1^2|^2+\frac{1}{4}|\alpha_3|^4
+\frac{1}{2}|\alpha_1\alpha_3|^2
=\frac{1}{2}+\frac{\alpha_1^2+(\alpha_1^*)^2}{4}
\label{twophoton}
\end{eqnarray}

This formula of double count $P_{UU}$ is same as that of single photon
interference $P_L$ except that
$\alpha_1$ is replaced by its square $\alpha_1^2$, and that the amplitude is
one half.. The half-period oscillation arises from the product $\alpha_1^2$
of the correlation function $\alpha_1$.

\subsection{Transform limited Gaussian photons}

To visualize the difference between one photon (or classical) 
interference and the two photon interference, let us calculate
probabilities as a function of displacement $\Delta\xi$
when the incoming photon has the transform limited Gaussian shape, 
\begin{equation}
E_1(\xi)=\exp\biggl(-\xi^2/(2\xi_0^2)+ik_0\xi\biggr).
\label{gaussian}
\end{equation}

The correlation function is
\begin{equation}
\alpha_1=\int E_1(\xi+\Delta\xi)E_1^*(\xi) d\xi
=\exp\biggl(-(\Delta\xi)^2/(4\xi_0^2)+ik_0\Delta\xi\biggr)
\label{alpha}
\end{equation}

Inserting this equation into Eq.(\ref{onephoton}),
 the single-photon interference pattern is
\begin{eqnarray}
\nonumber
P_U=\frac{1}{2}\biggl\{1-\exp\biggl(-(\Delta\xi)^2/(4\xi_0^2)\biggr)
\cos(k_0\Delta\xi)\biggr\},
\\
P_L=\frac{1}{2}\biggl\{1+\exp\biggl(-(\Delta\xi)^2/(4\xi_0^2)\biggr)
\cos(k_0\Delta\xi)\biggr\},
\label{gausspc}
\end{eqnarray}
whereas the two-photon interference is from Eq.~(\ref{twophoton}),
\begin{equation}
P_{UU}=P_{LL}=\frac{1}{4}\biggl\{1-\exp\biggl(-(\Delta\xi)^2/(2\xi_0^2)\biggr)
\cos(2k_0\Delta\xi)\biggr\}.
\label{gausspcc}
\end{equation}

\begin{figure}[htbp]
\begin{center}
\includegraphics[width=14cm]{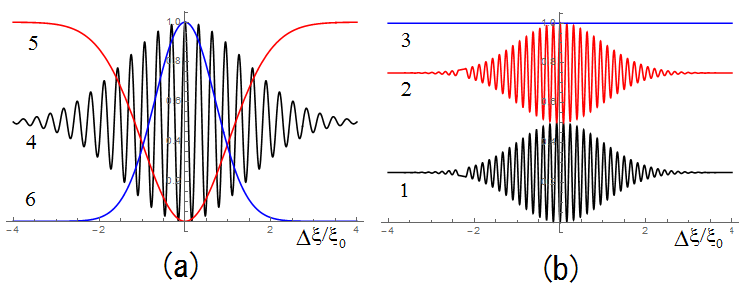}
\caption{(a): 4(black): single photon interference $P_U$ (Eq..~(\ref{gausspc}))
for $k_0\xi_0=10$,
5(red): Hong-Ou-Mandel's coincidence rate, 6(blue):
intensity profile of the input photon. 
(b): 1(black): coincidence probability $P_{UU}$
(Eq..~(\ref{gausspcc})) for $k_0\xi_0=10$, 2(red): probability when the detector 
at the position D detects one or two photons $P_{UU}+P_{UL}$,
and 3(blue): number of photons expected to detect at the position D per event 
$2P_{UU}+P_{UL}$.  }
\label{k2k}
\end{center}
\end{figure}

The patterns of $P_U$ of Eq.~(\ref{gausspc}) and $P_{UU}$ of 
Eq.~(\ref{gausspcc}) are shown in Fig.~\ref{k2k}(a-4) and (b-1), respectively.
Two photon interference oscillates twice faster, but its amplitude is
1/2.
The width of the oscillating pattern is  
$\sqrt{2}$ times narrower than the single photon interference 
Eq.~(\ref{onephoton}). This is natural because $\alpha_1^2$ is
expected to be $\sqrt{2}$ times
sharper than $\alpha_1$. Oscillation disappears completely at larger 
$\Delta\xi$, where the waves in U and L do not overlap at the beam splitter
BS$_2$.

The observed signal depends on the characteristics of the detector.
If the output of the channel U detector is proportional to the number 
of photons, the detected signal is constant
$2P_{UU}+P_{UD}=1$ as shown in curve 3 of Fig.~\ref{k2k}(b).
If the detector clicks only once regardless of the number of photons,
the output is proportional to $P_{UU}+P_{UL}$. This has the same shape
as $P_{UU}$, but is accompanied by  the constant background of 1/2. 
Figure~\ref{k2k}(a) shows also intensity profile of the input pulse 
(6:blue curve) and Hong-Ou-Mandel's dip (5: red curve).

 \subsection{Mach-Zhender interferometer: with a delay line and a phase 
 shifter}\label{inputshift}
 
 The  half-period oscillation pattern is considerably different,
 if the delay line is placed between L$_Z$ and L$_A$
 to change $\Delta\xi$, and the second delay line
 between L$_B$ and L$_C$ is used only to measure the amplitude of the
 half-period oscillation.  Then,
\begin{eqnarray}
\nonumber
f_{1,U_Z}^{\dagger}f_{1,L_Z}^{\dagger}
=f_{1,U_A}^{\dagger}f_{2,A_L}^{\dagger}
=f_{1,A_U}^{\dagger}(\alpha_1f_{1,L_A}^{\dagger}
+\alpha_3f_{3,L_A}^{\dagger})
\\
\nonumber
=\frac{1}{2}(f_{1,U_B}^{\dagger}+f_{1,L_B}^{\dagger})
(\alpha_1f_{1,U_B}^{\dagger}+\alpha_3f_{3,U_B}^{\dagger}
-\alpha_1f_{1,L_B}^{\dagger}-\alpha_3f_{3,L_B}^{\dagger})
\\
\nonumber
=\frac{1}{2}(f_{1,U_C}^{\dagger}+\mbox{e}^{i\zeta}f_{1,L_C}^{\dagger})
(\alpha_1f_{1,U_C}^{\dagger}+\alpha_3f_{3,U_C}^{\dagger}
-\alpha_1\mbox{e}^{i\zeta}f_{1,L_C}^{\dagger}
-\alpha_3\mbox{e}^{i\zeta}f_{3,L_C}^{\dagger})
\\
\nonumber
=\frac{1}{4}\biggl[\biggl\{(1+\mbox{e}^{i\zeta})f_{1,U_D}^{\dagger}
+(1-\mbox{e}^{i\zeta})f_{1,L_D}^{\dagger}\biggr\}
\\
\nonumber
\biggl\{\alpha_1(1-\mbox{e}^{i\zeta})f_{1,U_D}^{\dagger}
+\alpha_1(1+\mbox{e}^{i\zeta})f_{1,L_D}^{\dagger}
+\alpha_3(1-\mbox{e}^{i\zeta})f_{3,U_D}^{\dagger}
+\alpha_3(1+\mbox{e}^{i\zeta})f_{3,L_D}^{\dagger}\biggr\}\biggr]
\\
\nonumber
=\frac{1}{4}\biggl\{\sqrt{2}\alpha_1(1-\mbox{e}^{2i\zeta})
\frac{(f_{1,U_D}^{\dagger})^2}{\sqrt{2}}
+\alpha_3(1-\mbox{e}^{2i\zeta})f_{1,U_D}^{\dagger}f_{3,U_D}^{\dagger}
\\
\nonumber
+\sqrt{2}\alpha_1(1-\mbox{e}^{2i\zeta})
\frac{(f_{1,L_D}^{\dagger})^2}{\sqrt{2}}
+\alpha_3(1-\mbox{e}^{2i\zeta})f_{1,L_D}^{\dagger}f_{3,L_D}^{\dagger}
\\
+2\alpha_1(1+\mbox{e}^{2i\zeta})f_{1,U_D}^{\dagger}f_{1,L_D}^{\dagger}
+\alpha_3(1+\mbox{e}^{i\zeta})^2f_{1,U_D}^{\dagger}f_{3,L_D}^{\dagger}
+\alpha_3(1-\mbox{e}^{i\zeta})^2f_{3,U_D}^{\dagger}f_{1,L_D}^{\dagger}\biggr\},
\label{MZtwophoton}
\end{eqnarray}
where $\zeta=k\Delta\xi_{\zeta}$, and $\Delta\xi_{\zeta}$ is the path length 
shift of the delay line DL$_2$. 
$\Delta\xi_{\zeta}$ is assumed to be much smaller than the coherent length 
of the photon. 
Then, the probabilities are
\begin{eqnarray}
P_{UU}=P_{LL}=\frac{1}{16}(2|\alpha_1|^2+|\alpha_3|^2)|1-\mbox{e}^{2i\zeta}|^2
=\frac{1}{4}(1+|\alpha_1|^2)\sin^2\zeta
\\
\nonumber
P_{UL}=\frac{1}{16}\biggl\{4|\alpha_1|^2|1+\mbox{e}^{2i\zeta}|^2
+|\alpha_3|^2(|1+\mbox{e}^{i\zeta}|^4+|1-\mbox{e}^{i\zeta}|^4)\biggr\}
\\
=\frac{1}{2}\biggl\{(1+|\alpha_1|^2)\cos^2\zeta+|\alpha_3|^2\biggr\}
\end{eqnarray}

\begin{figure}[htbp]
\begin{center}
\includegraphics[width=8cm]{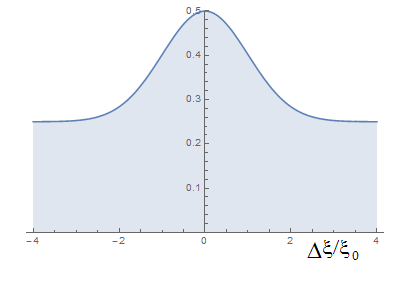}
\caption{The double-count rate $P_{UU}$ in Eq.~(\ref{twoshifters})
as a function of the delay length $\Delta\xi$ of the first delay line DL$_1$.
The shaded area shows the amplitude of the half-period oscillation
observable when the second delay line DL$_2$ is scanned..
There is no constant background. }
\label{twophoton6}
\end{center}
\end{figure}

When the incoming two photons have transform limited Gaussian shape Eq.~(\ref{gaussian}), the probability having double count is
\begin{equation}
P_{UU}=P_{LL}
=\frac{1}{4}\biggl(1+\mbox{e}^{-(\Delta\xi)^2/(2\xi_0^2)}\biggr)\sin^2\zeta,
\label{twoshifters}
\end{equation}
where $\Delta\xi$ is the path length shift of the DL$_1$.
The amplitude of the half-period oscillation remains one half of the peak value 
even when 
$\Delta\xi$ is large, and $f_{1,U_A}^{\dagger}|0>$ is completely separated 
from $f_{2,L_A}^{\dagger}|0>$ (Fig.~\ref{twophoton6}). 
In the present configuration the wave exists 
in the U and L channels simultaneously
at any time at some $\xi$. Therefore, the interferometric oscillation 
can always exists.

When two pulses arrive the point A at separate time, 
the process is equivalent to the 
sequence of two independent single-photon events.
The first event is single-photon interference of $E_1$ 
with the photon entering through
U$_A$. The second event is single-photon interference of $E_3$ 
with the photon 
entering through L$_A$.
Then, $P_{UU}$ of Eq.~(\ref{twoshifters}) with $\Delta\xi>>\xi_0$ 
\begin{equation}
P_{UU}=\frac{1}{4}\sin^2\zeta
\label{zeta}
\end{equation}
must be equal to the joint probability of finding a photon at U$_D$
when single photon enters from U$_A$,
 and finding a photon at U$_D$
when single photon enters from L$_A$. $P_U$ and $P_L$ in this case is
obtained by inserting $\alpha_1=\exp(i\zeta)$ in Eq.~(\ref{onephoton}).
 Since $P_U$ and $P_L$ exchanges 
when the input channel of the photon is reversed, the joint probability is
\begin{equation}
P_UP_L=\frac{1}{4}\biggl(1+\frac{\mbox{e}^{i\zeta}+\mbox{e}^{-i\zeta}}{2}\biggr)
\biggl(1-\frac{\mbox{e}^{i\zeta}+\mbox{e}^{-i\zeta}}{2}\biggr)
=\frac{1}{4}(1-\cos^2\zeta)=\frac{1}{4}\sin^2\zeta,
\end{equation}
which is equal to Eq.~(\ref{zeta}).

The appearance of sub-period oscillation is nothing to do with quantum
nature. The joint probability of two independent events 
is the product of
the probability of each event. If  each event oscillates at $k$,
the joint probability automatically generates the term which 
oscillates at $2k$ together with the term at $k$. When the latter disappears,
we observe the $2k$ periodicity. 

 Similar calssical sub-period oscillation can be observed 
 even when two photons overlap, 
 if $f_1^{\dagger}|0>$ and $f_2^{\dagger}|0>$ are orthogonal. 
 The state of two photons in the channel U at D is
 \[
 \frac{1}{4}\biggl(1-\mbox{e}^{2i\zeta}\biggr)f_{1,U_D}^{\dagger}
 f_{2,U_D}^{\dagger}
 \]
 This produces the probability
 \begin{equation}
 P_{UU}=\frac{1}{8}\biggl(1-\cos(2\zeta)\biggr).
 \end{equation}
 When two photons are identical, we obtain the same expression,
  \[
 \frac{1}{4}\biggl(1-\mbox{e}^{2i\zeta}\biggr)(f_{1,U_D}^{\dagger})^2.
 \]
 However, the probability is twice of the orthogonal case,
 \begin{equation}
 P_{UU}=\frac{1}{4}\biggl(1-\cos(2\zeta)\biggr),
 \end{equation}
 because of the quantum nature of the identical photons.
 
 The above discussion is valid only when $\Delta\xi\approx 0$,
 or $\Delta\xi$ is larger than the pulse length. We will show 
 an example of the intermediate case in the last section.

\subsection{$n$ photon beam splitter}

Comparison between Eq.~(\ref{onephoton}) and (\ref{twophoton}) shows
that the probability finding all atoms in single channel is expressed by an 
identical form with the parameter $\alpha_1$ replaced by $\alpha_1^2$.
This suggests that the result can be generalized to the $n$-photon case. 
Suppose that 
the first beam splitter is replaced by
a fictitious $n$-photon beam splitter which splits a single-mode $n$-photon
into two single-mode $n$-photons with the amplitude of $1/\sqrt{2}$,
\begin{eqnarray}
\nonumber
\frac{1}{\sqrt{n!}}(f_{1,U_A}^{\dagger})^n
=\frac{1}{\sqrt{2n!}}\biggl\{(f_{1,U_B}^{\dagger})^n\pm (f_{1,L_B}^{\dagger})^n
\biggr\}
\\
\frac{1}{2^{n+1}\sqrt{n!}}\biggl\{(f_{1,U_D}^{\dagger}+f_{1,L_D}^{\dagger})^n\pm
(\alpha_1f_{1,U_D}^{\dagger}-\alpha_1f_{1,L_D}^{\dagger}
+\alpha_3f_{3,U_D}^{\dagger}-\alpha_3f_{3,L_D}^{\dagger})^n\biggr\},
\end{eqnarray}
After expanding the above equation in terms of $f_{U_D}^{\dagger}$ and
$f_{L_D}^{\dagger}$, the terms with $n$-th power of 
$f_{U_D}^{\dagger}$'s are
\[
\frac{1}{\sqrt{2^{n+1}n!}}\biggl\{\sqrt{n!}(1\pm\alpha_1^n)
\frac{(f_{1,{U_D}}^{\dagger})^n}{\sqrt{n!}} \pm
\sum_{l=0}^{n-1}\frac{n!}{\sqrt{l!}\sqrt{(n-l)!}}
\alpha_1^l\alpha_3^{n-l}\frac{(f_{1,U_D}^{\dagger})^l}{\sqrt{l!}}
\frac{(f_{3,U_D}^{\dagger})^{n-l}}{\sqrt{(n-l)!}}
\biggr\}.
\]
Therefore, the probability having $n$ photons in the channel $U$ is
\begin{equation}
P_{U^n}=\frac{1}{2^{n+1}}\biggl\{1\pm(\alpha_1^n+(\alpha_1^*)^n)
+\sum_{l=0}^n\frac{n!}{l!(n-l)!}|\alpha_1|^{2l}|\alpha_3|^{2(n-l)}\biggr\}
=\frac{1}{2^{n+1}}\biggl\{2\pm((\alpha_1)^n+(\alpha_1^*)^n)\biggr\}.
\end{equation}
For the Gaussian pulse of Eq.(\ref{gaussian})
\begin{equation}
P_{U^n}=\frac{1}{2^n}\biggl\{1\pm\exp\biggl(-n(\Delta\xi)^2/(4\xi_0^2)\biggr)
\cos(nk_0\Delta\xi)\biggr\}.
\end{equation}
We do obtain $1/n$-period oscillation pattern after the second linear
beam splitter BS$_2$. However, its magnitude
diminishes rapidly as $2^{-n}$. The $1/n$-period term appears as a result of
the multiplicativity of $n$-photon Hilbert space. However, since photons
do not have interactions to keep $n$ photons together, 
it is natural that the probability 
of finding $n$ photons decreases rapidly with $n$. It is doubtful
that the construction of $1/n$ period interference pattern using linear
optics has any technically practical merit for precision measurement.

\subsection{Arbitrary single-mode photons  to channel U}

We show in this subsection that, when an 
arbitrary single-mode photon is sent to one of two channels, a single-photon
detector placed in one output channel will record the interference pattern
of the classical wave.

From Eq.~(\ref{singleph}),
\begin{equation} 
\nonumber
f_{1,U_A}^{\dagger}
=\frac{1}{2}\biggl\{(1+\alpha_1)f_{1,U_D}^{\dagger}
+\alpha_3f_{3,U_D}^{\dagger}
+(1-\alpha_1)f_{1,L_D}^{\dagger}
-\alpha_3f_{3,L_D}^{\dagger}\biggr\}.
\end{equation}
Therefore, using the commutation relations Eq.~(\ref{commutation}),
\begin{eqnarray}
f_{1,U_D}f_{1,U_A}^{\dagger}=f_{1,U_A}^{\dagger}f_{1,U_D}
+\frac{1}{2}(1+\alpha_1)
\\
f_{3,U_D}f_{1,U_A}^{\dagger}=f_{3,U_A}^{\dagger}f_{1,U_D}+\frac{\alpha_3}{2}
\end{eqnarray}

Using above equations the number of photons detected by the detector 
placed at U$_D$ from a $n$-photon state is
\begin{eqnarray}
\nonumber
<0|\frac{(f_{1,U_A})^n}{\sqrt{n!}}
 (f_{1,U_D}^{\dagger}f_{1,U_D}+f_{3,U_D}^{\dagger}f_{3,U_D})
 \frac{(f_{1,U_A}^{\dagger})^n}{\sqrt{n!}}|0>
 \\
 =\frac{n}{4}(1+\alpha_1)(1+\alpha^*_1) +\frac{n}{4}\alpha_3\alpha^*_3
 =\frac{n}{2}+\frac{n}{4}(\alpha_1+\alpha^*_1).
 \end{eqnarray}
Therefore, for any single mode photon states, 
\begin{equation}
|\Psi>\equiv\sum_n a_n|n>\equiv\sum_n a_n\frac{(f_{1,UA}^{\dagger})^n}{\sqrt{n!}}|0>
\end{equation}
we observe interference pattern which is same as that of a classical wave.
\begin{equation}
I_U=<\Psi|f_{1,U_D}^{\dagger}f_{1,U_D}+f_{3,U_D}^{\dagger}f_{3,U_D}|\Psi>
=\biggl\{\frac{1}{2}+\frac{1}{4}
(\alpha_1+\alpha^*_1)\biggr\}\sum_n n|a_n|^2.
\end{equation}
Similarly for the detector which is placed in the lower channel,
\begin{equation}
I_L=<\Psi|f_{1,L_D}^{\dagger}f_{1,L_D}+f_{3,L_D}^{\dagger}f_{3,L_D}|\Psi>
=\biggl\{\frac{1}{2}-\frac{1}{4}
(\alpha_1+\alpha^*_1)\biggr\}\sum_n n|a_n|^2.
\end{equation}

The above result verifies when the output of the interferometer
is detected by a photon-number detector, arbitrary single-mode 
photons fed through one channel 
will show the same interference pattern as that of a classical
interferometer. It also tells that for a small $\Delta\xi$, the pattern
is same for any input photon state.

This does not mean that the state $|\Psi>$ does not have
fractional-period oscillating terms. 
A simplest counter-example is the coincidence probability
of the $n=2$ Fock state. It is
\begin{equation}
P_{UL}=\frac{1}{4}\{1-\cos(2k_0\Delta\xi)\}.
\end{equation}
We need two detectors placed in U and L channels and a coincidence
electronics. To observe $1/n$-period oscillation
we need a detector system which can detect exclusively 
$n$ photon states,

\subsection{Homodyne detection}

Homodyne and heterodyne detections are the technique to measure the
amplitude of electro-magnetic waves. The standard technique is
interferometric measurement, where the reference wave is mixed with
the investigating wave. Since structure of the quantum electro-magnetic
wave is not same as the classical electro-magnetic field, it is not 
obvious if the interferometric measurement really measure the field
amplitude.
In the following we discuss the dynamics when all input photons are in the 
same mode.

Suppose $F_U(x)$ and $F_L(x)$ are polynomial functions of $x$ with
unity norm. We send photons $F_U(f_{U_Z}^{\dagger})$ and 
$F_L(f_{L_Z}^{\dagger})$ through point Z. The standard homodyne
detection consists of two photon-number detectors at B and 
calculates the difference of the output of two detectors.
\begin{eqnarray}
\nonumber
I_{hom}=<0|F_L^*(f_{L_Z})F_U^*(f_{U_Z})
(f_{U_B}^{\dagger}f_{U_B}-f_{L_B}^{\dagger}f_{L_B})
F_U(f_{U_Z}^{\dagger})F_L(f_{L_Z}^{\dagger})|0>
\\
\nonumber
=<0|F_L^*(\mbox{e}^{-i\phi}f_{L_A})F_U^*(f_{U_A})
(f_{U_B}^{\dagger}f_{U_B}-f_{L_B}^{\dagger}f_{L_B})
F_U(f_{U_A}^{\dagger})F_L(\mbox{e}^{i\phi}f_{L_A}^{\dagger})|0>
\\
\nonumber
=<0|F^*_L(\mbox{e}^{-i\phi}\frac{f_{U_B}-f_{L_B}}{\sqrt{2}})
F^*_U(\frac{f_{U_B}+f_{L_B}}{\sqrt{2}})
\\
\nonumber
(f_{U_B}^{\dagger}f_{U_B}-f_{L_B}^{\dagger}f_{L_B})
F_U(\frac{f_{U_B}^{\dagger}+f_{L_B}^{\dagger}}{\sqrt{2}})
F_L(\mbox{e}^{i\phi}\frac{f_{U_B}^{\dagger}-f_{L_B}^{\dagger}}{\sqrt{2}})|0>
\\
\nonumber
=<0|F^*_L(\mbox{e}^{-i\phi}\frac{f_{U_B}-f_{L_B}}{\sqrt{2}})
F^*_U(\frac{f_{U_B}+f_{L_B}}{\sqrt{2}})
\\
\nonumber
\frac{1}{2}\biggl\{(f_{U_B}^{\dagger}+f_{L_B}^{\dagger})(f_{U_B}-f_{L_B})
+(f_{U_B}^{\dagger}-f_{L_B}^{\dagger})(f_{U_B}+f_{L_B})\biggr\}
\\
F_U(\frac{f_{U_B}^{\dagger}+f_{L_B}^{\dagger}}{\sqrt{2}})
F_L(\mbox{e}^{i\phi}\frac{f_{U_B}^{\dagger}-f_{L_B}^{\dagger}}{\sqrt{2}})|0>,
\label{homodyne}
\end{eqnarray}  
where $\phi=k_0\Delta\xi$ is the phase shift caused by the delay line
DL$_1$, in which we assumed $\Delta\xi$ is small.
When  
$F_L(f^{\dagger})|0>$ is the coherent state\cite{Glauber} 
with eigenvalue $\alpha$,
then,
\begin{equation}
fF_L(f^{\dagger})|0>=\alpha F_L(f^{\dagger})|0>.
\end{equation}
Equation~(\ref{homodyne}) is written,
\begin{eqnarray}
\nonumber
I_{hom}=<0|F^*_L(\mbox{e}^{-i\phi}\frac{f_{U_B}-f_{L_B}}{\sqrt{2}})
F^*_U(\frac{f_{U_B}+f_{L_B}}{\sqrt{2}})
\\
\nonumber
\biggl\{\frac{\mbox{e}^{-i\phi}\alpha}{\sqrt{2}}
(f_{U_B}^{\dagger}+f_{L_B}^{\dagger})
+\frac{\mbox{e}^{i\phi}\alpha^*}{\sqrt{2}}(f_{U_B}+f_{L_B})\biggr\}
\\
\nonumber
F_U(\frac{f_{U_B}^{\dagger}+f_{L_B}^{\dagger}}{\sqrt{2}})
F_L(\mbox{e}^{i\phi}\frac{f_{U_B}^{\dagger}-f_{L_B}^{\dagger}}{\sqrt{2}})|0>
\\
=<0|F^*_U(f)(\mbox{e}^{-i\phi}\alpha f^{\dagger}
+\mbox{e}^{i\phi}\alpha^*f)F_U(f^{\dagger})|0>,
\label{homodyne2}
\end{eqnarray}
which is proportional to the homodyne signal of 
a single mode photon state $F_U(f^{\dagger})$. The last line of 
Eq.~(\ref{homodyne2}) is derived by using  
$[(f_{U_B}^{\dagger} \pm f_{L_B}^{\dagger}),(f_{U_B}\mp f_{L_B})]=0$. 
The output signal increases proportional to $|\alpha|$. However, this
does not guarantee the improvement in actual experiment, because
$f_{U_B}^{\dagger}f_{U_B}$ and $f_{L_B}^{\dagger}f_{L_B}$ are detected by
separate detectors. They contain the term proportional to $|\alpha|^2$,
and may generate technical noise.

\section{Interference with arbitrary photons}\label{arbitraryshape}

We try to derive general formulae to express interference
patterns for arbitrary input photons. For this purpose we use 
commonly used
formulation, expansion of photon state by monochromatic vectors. 
Then, the function
of a delay line is to multiply a phase factor on each vector, and 
do not change the wave form of the vector.

\subsection{Hong-Ou-Mandel's dip}

Consider when we place detectors at the position B and send 
arbitrary two photons from the position Z to measure the
Hong-Ou-Mandel's dip,
 
The photon state is
\begin{eqnarray}
\nonumber
\Psi_{\mbox{m}}=
\biggl(\sum_{j=-\infty}^{\infty}\beta_j f_{j,U_Z}^{\dagger}\biggr)
\biggl(\sum_{l=-\infty}^{\infty}\gamma_l f_{l,L_Z}^{\dagger}\biggr)=
\biggl(\sum_{j=-\infty}^{\infty}\beta_j f_{j,U_A}^{\dagger}\biggr)
\biggl(\sum_{l=-\infty}^{\infty}\gamma_l
\mbox{e}^{i\phi_l} f_{l,L_A}^{\dagger}\biggr)
\\
=\frac{1}{2}
\sum_{i,j=-\infty}^{\infty}\beta_j\gamma_l\mbox{e}^{i\phi_l}
\biggl\{f_{j,U_B}^{\dagger}f_{l,U_B}^{\dagger}
-f_{j,L_B}^{\dagger}f_{l,L_B}^{\dagger}
-f_{j,U_B}^{\dagger}f_{l,L_B}^{\dagger}
+f_{j,L_B}^{\dagger}f_{l,U_B}^{\dagger}\biggr\}
\label{mandelraw}
\end{eqnarray}
Rewriting above expression in terms of orthonormal vectors,
\begin{eqnarray}
\nonumber
\Psi_{\mbox{m}}=
\frac{1}{2}\biggl\{\sum_{j<l}(\beta_j\gamma_l\mbox{e}^{i\phi_l}
+\beta_l\gamma_j\mbox{e}^{i\phi_j})f_{j,U_B}^{\dagger}f_{l,U_B}^{\dagger}
+\sum_j\sqrt{2}\beta_j\gamma_j\mbox{e}^{i\phi_j}(\frac{(f_{j,U_B}^{\dagger})^2}
{\sqrt{2}}\biggr\}
\\
\nonumber
-\frac{1}{2}\biggl\{\sum_{j<l}(\beta_j\gamma_l\mbox{e}^{i\phi_l}
+\beta_l\gamma_j\mbox{e}^{i\phi_j})f_{j,L_B}^{\dagger}f_{l,L_B}^{\dagger}
+\sum_j\sqrt{2}\beta_j\gamma_j\mbox{e}^{i\phi_j}
\frac{(f_{j,L_B}^{\dagger})^2}{\sqrt{2}})\biggr\}
\\
+\frac{1}{2}\biggl\{\sum_{j,l}(\beta_l\gamma_j\mbox{e}^{i\phi_j}
-\beta_j\gamma_l\mbox{e}^{i\phi_l})f_{j,U_B}^{\dagger}f_{l,L_B}^{\dagger}
\biggr\}.
\label{appendix3}
\end{eqnarray}
The probability having one photon par channel, and two photons in
either channel are
\begin{eqnarray}
\nonumber
P_{UL_B}=\frac{1}{4}\sum_{j,l}w_jw_l|\beta_l\gamma_j\mbox{e}^{i\phi_j}-
\beta_j\gamma_l\mbox{e}^{i\phi_l}|^2=I_{a_m}-I_{c_m}
\\
P_{UU_B}=P_{LL_B}=
\frac{1}{8}\sum_{j,l}w_jw_l|\beta_l\gamma_j\mbox{e}^{i\phi_j}+
\beta_j\gamma_l\mbox{e}^{i\phi_l}|^2=\frac{1}{2}(I_{a_m}+I_{c_m})
\label{mandelprobe}
\end{eqnarray}
where
\begin{eqnarray}
I_{a_m}=\frac{1}{2}\sum_{j,l}w_jw_l
|\beta_j|^2|\gamma_l|^2
\label{m1}
\\
\nonumber
I_{c_m}=\frac{1}{4}\sum_{j,l}w_jw_l
(\beta_l\gamma_l^*\beta_j^*\gamma_j
\mbox{e}^{i\phi_j-i\phi_l}+\beta_l^*\gamma_l\beta_j\gamma_j^*
\mbox{e}^{i\phi_l-i\phi_j})
\\
=\frac{1}{2}\biggl\{\sum_j\omega_j\beta_j\gamma^*_j\mbox{e}^{i\phi_j}\biggr\}
\biggl\{\sum_j\omega_j\beta^*_j\gamma_j\mbox{e}^{-i\phi_j}\biggr\}
\label{m2}
\end{eqnarray}

In integral form the above expressions are,
\begin{eqnarray}
I_{a_m}=\frac{1}{2}\biggl\{\int_{-\infty}^{\infty}dk
w(k)|\beta(k)|^2\biggr\}
\biggl\{\int_{-\infty}^{\infty}dk^{\prime}
w(k^{\prime})|\gamma(k^{\prime})|^2\biggr\}
\label{mandelconst}
\\
I_{c_m}=\frac{1}{2}\biggl\{\int_{-\infty}^{\infty}dk
w(k)\beta^*(k)\gamma(k)\mbox{e}^{-i\phi(k)}\biggr\}
\biggl\{\int_{-\infty}^{\infty}dk^{\prime}
w(k^{\prime})\beta(k^{\prime})\gamma^*(k^{\prime})
\mbox{e}^{i\phi(k^{\prime})}\biggr\},
\label{mandeldip}
\end{eqnarray}
where $\phi(k)=(k_0+k)\Delta\xi$.

We divided $P_{UL_B}$, $P_{UU_B}$, and $P_{LL_B}$ into sum of two terms,
$I_{a_m}$ and $I_{c_m}$, whose dynamics are fairly different.
$I_{a_m}$ does not depend on the phase of the Fourier transform $\beta(k)$
and $\gamma(k)$, and is constant 1/2, if $w(k)=1$.

\begin{figure}[htbp]
\begin{center}
\includegraphics[width=16cm]{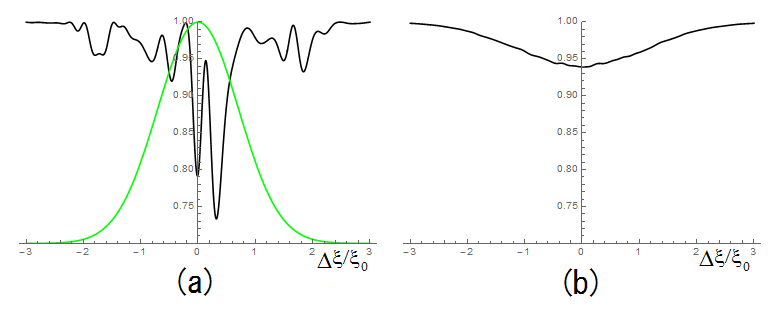}
\caption{Hong-Ou-Mandel's spectrum  for the photons with Gaussian
envelope and random phase. See Ref.\cite{simu} for the exact pulse shape.
(a): pattern of single event. (b) pattern after 5000 random input events.
Green curve is the intensity profile of the input photon.}
\label{MZm}
\end{center}
\end{figure}

$I_{c_m}$ resuts from the cross terms of Eq.~(\ref{appendix3}) and
depends on the phase of $\beta(k)\gamma^*(k^{\prime})$.
If $\beta(k)$
and $\gamma(k)$ are not correlated, the phase of the factors
$\beta(k)\gamma^*(k)$ and $\beta^*(k^{\prime})\gamma(k^{\prime})$
changes randomly.  Since $\beta(k)$ is the Fourier
transform of the input wave, $I_{c_m}\rightarrow 0$ 
for $\xi_0 \rightarrow\infty$. However, if the pulse length $\xi_0$ is 
finite, $\beta(k)$ does not change grossly
up to $k\pm \Delta k$, where $\Delta k=\pi/\xi_0$.
This situation is same for $\gamma(k)$.
Therefore, the integral in Eq.~(\ref{mandeldip}) has non-zero value of
roughly $(1/k_{\mbox{max}})\sqrt{k_{\mbox{max}}/\Delta k}$
up to $\Delta\xi\leq\Delta\xi_{\mbox{pulse}}$,
where $k_{\mbox{max}}$ is the full width of $\beta(k)$, or equivalently
the coherence width of the input pulse $\xi_{\mbox{coh}}$.
Therefore, $I_{c_m}(\Delta\xi)\sim \xi_{\mbox{coh}}/
\xi_0$ for $\Delta\xi\leq\xi_0$. 
The coincidence probability $P_{UL_B}$ is $I_{a_m}-I_{c_m}$. This produces the
Mandel's dip of the width roughly the input pulse length $2\xi_0$.
The depth of the dip
decreases as the coherence length of the photon decreases.
Real shape of the dip can be complicated reflecting the phase variation 
of the photon pulse as shown in Fig.~\ref{MZm}(a). However, the dip 
does not disappear, even after the signal is averaged over many events 
(Fig.~\ref{MZm}(b)), because $I_{c_m}$ is a positive-definite function. 
Therefore, the gross shape of the dip
 is a good measure of the photon's pulse length.
The bottom of the dip reaches zero only when two input photons 
have the same shape as it is well known from literatures.

If the sensitivity of the detector is limited to a very narrow spectral range,
it is equivalent to reduce $k_{\mbox{max}}$ accordingly.
The relative depth of the dip does not change,
 but its absolute magnitude decreases.

\subsection{Two photon Mach-Zhender}\label{TMZ}

Let us consider the case, when two arbitrary photons enter the two
channels of the interferometer. 

\begin{eqnarray}
\nonumber
\Psi_{\mbox{d}}=
\biggl(\sum_{j=-\infty}^{\infty}\beta_j f_{j,U_A}^{\dagger}\biggr)
\biggl(\sum_{l=-\infty}^{\infty}\gamma_l f_{l,L_A}^{\dagger}\biggr)
\\
\nonumber
=\frac{1}{2}\biggl\{\sum_j\beta_j(f_{j,U_B}^{\dagger}+f_{j,L_B}^{\dagger})
\sum_l\gamma_l(f_{l,U_B}^{\dagger}-f_{l,L_B}^{\dagger})\biggr\}
\\
\nonumber
=\frac{1}{2}\biggl\{\sum_j\beta_j(f_{j,U_C}^{\dagger}+
\mbox{e}^{i\phi_j}f_{j,L_C}^{\dagger})
\sum_l\gamma_l(f_{l,U_C}^{\dagger}-
\mbox{e}^{i\phi_l}f_{l,L_C}^{\dagger})\biggr\}
\\
\nonumber
=\frac{1}{4}\biggl[\sum_j\beta_j\biggl\{(1+\mbox{e}^{i\phi_j})
f_{j,U_D}^{\dagger}+(1-\mbox{e}^{i\phi_j})f_{j,L_D}^{\dagger}\biggr\}
\sum_l\gamma_l\biggl\{(1-\mbox{e}^{i\phi_l})
f_{l,U_D}^{\dagger}+(1+\mbox{e}^{i\phi_l})f_{l,L_D}^{\dagger}\biggr\}\biggr]
\\
\nonumber
=\frac{1}{4}\sum_{j,l} \biggl\{\beta_j\gamma_l(1+\mbox{e}^{i\phi_j})
(1-\mbox{e}^{i\phi_l})f_{j,U_D}^{\dagger}f_{l,U_D}^{\dagger}
+\beta_j\gamma_l(1-\mbox{e}^{i\phi_j})
(1+\mbox{e}^{i\phi_l})f_{j,L_D}^{\dagger}f_{l,L_D}^{\dagger}
\\
(1+\mbox{e}^{i\phi_j})(1+\mbox{e}^{i\phi_l})\beta_j\gamma_lf_{j,U_D}^{\dagger}
f_{l,L_D}^{\dagger}+(1-\mbox{e}^{i\phi_j})(1-\mbox{e}^{i\phi_l})
\beta_j\gamma_lf_{j,L_D}^{\dagger}f_{l,U_D}^{\dagger}\biggr\}
\label{twophotonraw}
\end{eqnarray}
The above
expression contains terms with identical vectors such as 
$f_{j,U_D}^{\dagger}f_{l,U_D}^{\dagger}$ and the term with $j$ and $l$ 
exchanged. Furthermore, $f_{j,U_D}^{\dagger}f_{j,U_D}^{\dagger}|0>$ is
not a normalized vector. (see Eq.~(\ref{normalizedn}) )

The expression in terms of 
normalized orthonormal vectors is
\begin{eqnarray}
\nonumber
\Psi_{\mbox{d}}
=\frac{1}{4}\sum_{j<l}\biggl\{\beta_j\gamma_l(1+\mbox{e}^{i\phi_j})
(1-\mbox{e}^{i\phi_l})+\beta_l\gamma_j(1+\mbox{e}^{i\phi_l})
(1-\mbox{e}^{i\phi_j})\biggr\}f_{j,U_D}^{\dagger}f_{l,U_D}^{\dagger}
\\
\nonumber
+\frac{1}{2\sqrt{2}}\sum_j\beta_j\gamma_j(1-\mbox{e}^{2i\phi_j})
\frac{(f_{j,U_D}^{\dagger})^2}{\sqrt{2}}
+\frac{1}{2\sqrt{2}}\sum_j\beta_j\gamma_j(1-\mbox{e}^{2i\phi_j})
\frac{(f_{j,L_D}^{\dagger})^2}{\sqrt{2}}
\\
\nonumber
+\frac{1}{4}\sum_{j<l}\biggl\{\beta_j\gamma_l(1-\mbox{e}^{i\phi_j})
(1+\mbox{e}^{i\phi_l})+\beta_l\gamma_j(1-\mbox{e}^{i\phi_l})
(1+\mbox{e}^{i\phi_j})\biggr\}f_{j,L_D}^{\dagger}f_{l,L_D}^{\dagger}
\\
+\frac{1}{4}\sum_{j,l}\biggl\{\beta_j\gamma_l(1+\mbox{e}^{i\phi_j})
(1+\mbox{e}^{i\phi_l})+\beta_l\gamma_j(1-\mbox{e}^{i\phi_j})
(1-\mbox{e}^{i\phi_l})\biggr\}f_{j,L_D}^{\dagger}f_{l,U_D}.
\label{appendix2}
\end{eqnarray}

\noindent
The probability having two photons in channel U$_D$ is obtained by
summing intensity of the relevant terms in Eq.~(\ref{appendix2}).
\begin{eqnarray}
\nonumber
P_{UU_D}=
\frac{1}{16}\sum_{j<l}w_jw_l\biggl|\beta_j\gamma_l(1+\mbox{e}^{i\phi_j})
(1-\mbox{e}^{i\phi_l})+\beta_l\gamma_j(1+\mbox{e}^{i\phi_l})
(1-\mbox{e}^{i\phi_j})\biggr|^2
\\
\nonumber
+\frac{1}{8}\sum_jw_j^2\biggl|
\beta_j\gamma_j(1-\mbox{e}^{2i\phi_j})\biggr|^2
\\
=\frac{1}{32}\sum_{j,l}w_jw_l\biggl|\beta_j\gamma_l(1+\mbox{e}^{i\phi_j})
(1-\mbox{e}^{i\phi_l})+\beta_l\gamma_j(1+\mbox{e}^{i\phi_l})
(1-\mbox{e}^{i\phi_j})\biggr|^2=I_a+I_c,
\label{twocorrelated}
\end{eqnarray}
where
\begin{eqnarray}
I_a=\frac{1}{16}\sum_{j,l}\beta_j\beta^*_j\gamma_l\gamma^*_l
w(k)w(k^{\prime})(2+\mbox{e}^{i\phi_j}+\mbox{e}^{-i\phi_j})
(2-\mbox{e}^{i\phi_l}-\mbox{e}^{-i\phi_l}),
\label{Ia}
\\
I_c=\frac{1}{16}\sum_{j,l}\beta_j\gamma^*_j\beta^*_l\gamma_l
w(k)w(k^{\prime})(\mbox{e}^{i\phi_j}-\mbox{e}^{-i\phi_j})
(\mbox{e}^{-i\phi_l}-\mbox{e}^{i\phi_l}).
\label{Ic}
\end{eqnarray}
Expression in integral form is 
\begin{eqnarray}
\nonumber
I_a=\frac{1}{16}
\\
\biggl\{\int_{-\infty}^{\infty}dkw(k)
|\beta(k)|^2
(2+\mbox{e}^{i\phi(k)}+\mbox{e}^{-i\phi(k)})\biggr\}
\biggl\{\int_{-\infty}^{\infty}dk^{\prime}w(k^{\prime})
|\gamma(k^{\prime})|^2
(2-\mbox{e}^{i\phi(k^{\prime})}-\mbox{e}^{-i\phi(k^{\prime})})\biggr\},
\label{t1}
\\
\nonumber
I_c=\frac{1}{16}
\\
\biggl\{\int_{-\infty}^{\infty}dkw(k)\beta(k)\gamma^*(k)
(\mbox{e}^{i\phi(k)}-\mbox{e}^{-i\phi(k)})\biggr\}
\biggl\{\int_{-\infty}^{\infty}dk^{\prime}w(k^{\prime})
\beta^*(k^{\prime})\gamma(k^{\prime})
(\mbox{e}^{-i\phi(k^{\prime})}-\mbox{e}^{i\phi(k^{\prime})})\biggr\},
\label{t2}
\end{eqnarray}
where $\phi(k)=(k_0+k)\Delta\xi$.

Again we divided $P_{UU_D}$ 
into two terms, $I_a$ (Eq.~(\ref{Ia})) and 
$I_c$ (Eq.~(\ref{Ic})). 

$I_a$ does not depend on the phase variation
of $\beta(k)$ or $\gamma(k)$. It is composed of the product of two
integrated terms, each of which has a constant term and two terms
oscillating by $\exp(\pm k_0\Delta\xi)$. 
The constant term is 1/4. 
The oscillating terms has 
the magnitude of the order 
$\delta\equiv \xi_{\mbox{coh}}/\xi_0$,
because only the $\delta$ portion of the integral contributes to the
result. The classical $2\pi/k_0$-period oscillating term
arises from the product of the constant term and the oscillating term.
Therefore, its amplitude is in the order of
$\delta$. The half-period  $\pi/k_0$
oscillating terms arises from the product of two classical-oscillating terms, 
and its amplitude is $\delta^2$. Therefore, generally the classical
oscillating term dominates.

The $I_c$ is influenced by the phase variation of $\beta(k)$ and 
$\gamma(k)$. This term does not have a constant term of unity
magnitude. It has the half-period oscillating term and constant term
of magnitude $\delta^2$.

The broad oscillating terms
will be averaged to zero when the event is repeated
by photons of randomly varying phase. 

\begin{figure}[htbp]
\begin{center}
\includegraphics[width=16cm]{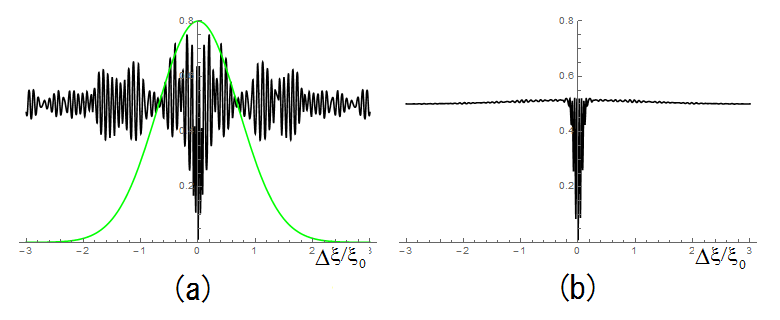}
\caption{Two-photon interference pattern 
for the same
Gaussian random phase input sequence in Fig.~\ref{MZm}.
(a): pattern of single event. (b) pattern after 5000 random input events.
Green curve is the intensity profile of the input photon.}
\label{MZd}
\end{center}
\end{figure}

Figure~\ref{MZd} shows the interference pattern for the same
sequence of input pulses as in the previous subsection.

\section{Linearly chirped photons, Gaussian envelope}\label{third}

When two photons are produced by the instantaneous parametric down 
conversion from the transform limited parent photon, 
the instantaneous frequency of generated two photons
are expected to be oppositely chirped with equal slope. One may expect that
the half-period oscillation has the same coherence length as that of  the parent photon of the parametric down conversion. It is not obvious if this
happens,
because all devices inside the interferometer acts on individual photon,
not simultaneously on two or more photons.

To check this assumption we calculate the interference pattern of
 oppositely-directed linearly-chirped Gaussian photons.
The result shows that the interference pattern is basically no difference from
that of general non-correlated photon pairs in the previous section.

\subsection{Pulse shape and detector response function}

The wave function of a linearly chirped Gaussian photon is
\begin{equation}
E(\xi)=\frac{1}{\pi^{1/4}\xi_0^{1/2}}
\exp\biggl(-\frac{\xi^2}{2\xi_0^2}-\frac{i\kappa\xi^2}{2\xi_0^2}+ik_0\xi\biggr),
\label{chirpedgauss}
\end{equation}
where $k_0$ is the center wave length.
Its Fourier transform with proper normalization is
\begin{equation}
\beta(k)=\frac{\xi_0^{1/2}}{\pi^{1/4}(1+\kappa^2)^{1/4}}
\exp\biggl(-\frac{\xi_0^2k^2}{2(1+i\kappa)}\biggr)
\label{beta}
\end{equation}
The second photon which is produced by degenerate parametric 
down conversion from a transform-limited Gaussian photon is
\begin{equation}
\gamma(k)=\beta^*(k)=\frac{\xi_0^{1/2}}{\pi^{1/4}(1+\kappa^2)^{1/4}}
\exp\biggl(-\frac{\xi_0^2 k^2}{2(1-i\kappa)}\biggr).
\label{gamma}
\end{equation}
We assum that the sensitivity function of the detector is
\begin{equation}
w(k)=\exp(-\eta\xi_0^2 k^2),
\end{equation}
or, in sum form
\begin{equation}
w_j=\exp\biggl\{-\eta\xi_0^2 (j\Delta k)^2\biggr\},
\end{equation}
where $\Delta k$ is the mesh size of $k$, when the equation is expressed in
sum form..

\noindent
Note that the phase shift through the delay line is
\begin{equation}
\phi_j=k_f\Delta\xi=(k_0+j\Delta k)\Delta\xi,
\end{equation}
or in integral form
\begin{equation}
\phi(k)=(k_0+k)\Delta\xi
\label{sens}
\end{equation}

\subsection{Hong-Ou-Mandel's dip}

For the chirped Gaussian photons, in which 
Eqs.~(\ref{beta}) and (\ref{gamma}) are satisfied,
\begin{eqnarray}
I_{a_m}=\frac{1}{2}\int_{-\infty}^{\infty}dk\int_{-\infty}^{\infty}dk^{\prime}
w(k)w(k^{\prime})|\beta(k)\beta^*(k^{\prime})|^2
=\frac{1}{2(1+\eta+\eta\kappa^2)},
\\
\nonumber
I_{c_m}=\frac{1}{2}\int_{-\infty}^{\infty}dk\int_{-\infty}^{\infty}dk^{\prime}
w(k)w(k^{\prime})\beta(k)^2\beta^*(k^{\prime})^2
\mbox{e}^{i(k-k^{\prime})\Delta\xi}
\\
=\frac{1}{2\sqrt{(1+\kappa^2)\{(1+\eta)^2+\eta^2\kappa^2\}}}\exp\biggl\{
-\frac{1+\eta+\eta\kappa^2}{2\{(1+\eta)^2+\eta^2\kappa^2\}}
\frac{\Delta\xi^2}{\xi_0^2}\biggr\},
\\
P_{UL_B}=I_{a_m}-I_{c_m}.
\end{eqnarray}
The first term is constant of the delay $\Delta\xi$. Mandel's dip
arises from the cross term $I_{c_m}$
When the detector's spectral range is unlimited ($\eta=0$),
the Mandel's dip has always the length of the photon pulse, though the
depth decreases as the coherence length of the photon decreases.

\subsection{Two photon Mach-Zhender}

Consider when Eqs.~(\ref{beta}) and (\ref{gamma}) are satisfied, 
or more relaxed condition $\gamma(k)=\beta^*(k)$ is satisfied.
Inserting $\gamma(k)=\beta^*(k)$ into Eqs.~(\ref{Ia}) and (\ref{Ic}),
\label{doublecount}
\begin{eqnarray}
\nonumber
I_{a_d}=
\frac{1}{16}\int_{-\infty}^{\infty}dk\int_{-\infty}^{\infty}dk^{\prime}\biggl\{
4w(k)|\beta(k)|^2w(k^{\prime})|\beta(k^{\prime})|^2
\\
\nonumber
-w(k)|\beta(k)|^2\mbox{e}^{ik\Delta\xi}w(k^{\prime})|\beta^*(k^{\prime})|^2
\mbox{e}^{ik^{\prime}\Delta\xi}\mbox{e}^{2ik_0\Delta\xi}
\\
\nonumber
-w(k)|\beta(k)|^2\mbox{e}^{-ik\Delta\xi}w(k^{\prime})|\beta^*(k^{\prime})|^2
\mbox{e}^{-ik^{\prime}\Delta\xi}\mbox{e}^{-2ik_0\Delta\xi}
\\
\nonumber
-w(k)|\beta(k)|^2\mbox{e}^{ik\Delta\xi}w(k^{\prime})|\beta^*(k^{\prime})|^2
\mbox{e}^{-ik^{\prime}\Delta\xi}
\\
-w(k)|\beta(k)|^2\mbox{e}^{-ik\Delta\xi}w(k^{\prime})|\beta^*(k^{\prime})|^2
\mbox{e}^{ik^{\prime}\Delta\xi}\biggr\}
\\
\nonumber
I_{c_d}=
\frac{1}{16}\int_{-\infty}^{\infty}dk\int_{-\infty}^{\infty}dk^{\prime}\biggl\{
-w(k)\beta(k)^2\mbox{e}^{ik\Delta\xi}w(k^{\prime})\beta^*(k^{\prime})^2
\mbox{e}^{ik^{\prime}\Delta\xi}\mbox{e}^{2ik_0\Delta\xi}
\\
\nonumber
-w(k)\beta(k)^2\mbox{e}^{-ik\Delta\xi}w(k^{\prime})\beta^*(k^{\prime})^2
\mbox{e}^{-ik^{\prime}\Delta\xi}\mbox{e}^{-2ik_0\Delta\xi}
\\
+w(k)\beta(k)^2\mbox{e}^{ik\Delta\xi}w(k^{\prime})\beta^*(k^{\prime})^2
\mbox{e}^{-ik^{\prime}\Delta\xi}
+w(k)\beta(k)^2\mbox{e}^{-ik\Delta\xi}w(k^{\prime})\beta^*(k^{\prime})^2
\mbox{e}^{ik^{\prime}\Delta\xi}\biggr\}
\end{eqnarray}
After integration we obtain
\begin{eqnarray}
I_{a_d}=\frac{1}{4(1+\eta+\eta\kappa^2)}
-\frac{1}{8(1+\eta+\eta\kappa^2)}\exp\biggl\{-\frac{(1+\kappa^2)\Delta\xi^2}
{2(1+\eta+\eta\kappa^2)\xi_0^2}\biggr\}\{1+\cos(2k_0\Delta\xi)\}
\\
I_{c_d}=\frac{1}{8\sqrt{(1+\kappa^2)\{(1+\eta)^2+\eta^2\kappa^2\}}}
\exp\biggl\{-\frac{(1+\eta+\eta\kappa^2)\Delta\xi^2}
{2\{(1+\eta)^2+\eta^2\kappa^2\}\xi_0^2}\biggr\}\{1-\cos(2k_0\Delta\xi)\}
\label{pattern}
\end{eqnarray}

It is easy to see from Eq.~(\ref{Ia}) 
that the classical $2\pi/k_0$-period oscillating term in $I_a$
vanishes when $\gamma(k)=\beta^*(k)$. As a result the 
half-period oscillation in $I_c$ is observable over the entire pulse 
length $\Delta\xi_{\mbox{pulse}}$. 

When the detector response is instantaneous ($\eta=0$), 
 The half-period oscillation part of $I_a$ has the width roughly
 equal to the coherence length $\Delta\xi_{\mbox{coh}}$ with
 a smaller residual over $\Delta\xi_{\mbox{pulse}}$. 
The peak-to-peak amplitude of the oscillation is one-quarter, 
which is same as the case
of the transform limited pulse of Eq.~(\ref{gausspcc}).
$I_a$ is biased by a constant of 1/4.

The cross term $I_c$ oscillates from the base line. It has the length
of $\sqrt{2}$ times of the input photon.
Its peak-to-peak amplitude is $\sqrt{1+\kappa^2}$ times smaller than
that of $I_a$.

\begin{figure}[htbp]
\begin{center}
\includegraphics[width=10cm]{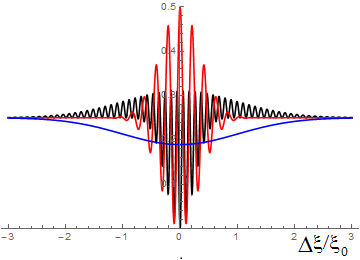}
\caption{Chirped pulse input. $\kappa=4$ and $\eta=0$. 
Black(half-period): two photon interference pattern $P_{UU_D}$, Red(single-period): one-half of the single photon interference pattern $P_{U_D}/2$, 
Blue(envelope): one-half of the Mandel's dip $P_{UL_B}/2$. 
Three curves are normalized to have the
same value at large $|\Delta\xi|$. The two-photon oscillation extends 
to the pulse width $\sim |\xi_0|$. However, its amplitude is small.
The main part has the same width as single-photon oscillation.}
\label{chirp}
\end{center}
\end{figure}

We show the single and two-photon interference patterns, and
Mandel's dip for $\kappa=4$ and $\eta=0$ in Fig.~\ref{chirp}.
The figure of $P_{UU_D}$ shows what we can expect from
this measurement. The half-period oscillation from the base line
is observed 
only around $\Delta\xi\approx 0$. Its width is $\sqrt{2}$ times the
coherence length of the input photon $\Delta\xi_{\mbox{coh}}$.
This width is $\sqrt{2}$ times narrower than the single-atom
interference.
It is produced from $I_{a_d}$. Broad oscillation of the half-period
covering the entire $\Delta\xi_{\mbox{pulse}}$
is added to the main contribution from $I_{c_d}$, but its amplitude is
much smaller than the main term.

When the detector's spectral response range is very narrow 
($\eta>\kappa^2$ and $\eta>1$), the interference pattern is the same
as that of the transform-limited Gaussian pulse Eq.(\ref{gausspcc}).
However, its absolute 
magnitude decreases, because the detector is sensitive
only to a small portion of the photon hitting the detector.

\section{Two orthogonal photons in Subsection~ \ref{inputshift}}

In the section \ref{inputshift} we showed that , 
when the photons enter U$_A$ and L$_A$ at separate time, the response of the interferometer is identical to that
of two independent single-photon interfereces. 
The same interpretation is possible if the two photons are not correlated 
at the first beam splitter BS$_1$ even when they overlap temporally. 
This is seen from the expressions Eq.~(\ref{t1}), 
and (\ref{t2}) of Sec.~\ref{TMZ} in the general input case.
Rewriting $\xi$ by $\xi_2$, and assuming $\Delta\xi_2 k_{max}<<\pi$,
where $k_{max}$ is the maximum extent of the wave vector of 
the photons, we get $\zeta=(k_0+k)\Delta\xi_2\approx k_0\Delta\xi_2$.
Then, Eq.~(\ref{t1}) and (\ref{t2}) are reduced to
\begin{eqnarray}
\nonumber
I_a=\frac{1}{4}\sin^2 (k_0\Delta\xi_2)
\\
I_c=\frac{1}{4}\biggl|\int_{\infty}^{\infty}dk\beta(k)\gamma^*(k)\biggr|^2
\sin^2(k_0\Delta\xi_2).
\label{small}
\end{eqnarray}  
where we assumed that the spectral response of the detector is flat
$w(k)=1$.
Both $I_a$ and $I_c$ oscillate at double frequency of the single photon
interference with the same phase.
If the input two photons do not have correlation, $I_c=0$,
then, the amplitude of half-period oscillation is 1/4, which is a half
of the peak amplitude of the identical two-photon case.

\begin{figure}[htbp]
\begin{center}
\includegraphics[width=10cm]{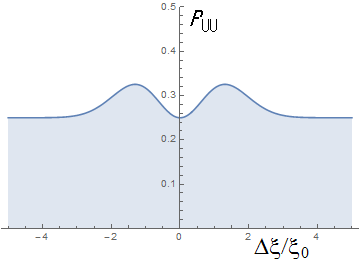}
\caption{Double count probability $P_{UU}$ for the orthogonal photons 
input.  The shaded area shows the amplitude of the half-period scillation.
The magnitude is 1/4 for large $\Delta\xi_1$ as well as
for $\Delta\xi_1=0$.}
\label{twoortho}
\end{center}
\end{figure}

We show in Fig.~\ref{twoortho} the double-count probability $P_{UU}$
for the two ortogonal Gaussian pulse input. We choose for the 
input photons 
\begin{eqnarray}
\nonumber
E_1(\xi)=\frac{1}{\pi^{1/4}\xi_0^{1/2}}\exp(-\frac{\xi^2}{2\xi_0^2}).
\\
E_2(\xi)=\frac{2^{1/2}}{\pi^{1/4}\xi_0^{1/2}
\{1-\exp(-\lambda^2\xi_0^2)\}^{1/2}}
\sin(\lambda\xi_0)\exp(-\frac{\xi^2}{2\xi_0^2}).
\end{eqnarray}
Then $\beta(k)$ and $\gamma(k)$ are
\begin{eqnarray}
\nonumber
\beta(k)=\frac{\xi_0^{1/2}}{\pi^{1/4}}
\exp(\frac{-\xi_0^2 k^2}{2}+ik\Delta\xi_1)
\\
\gamma(k)=\frac{\xi_0^{1/2}}
{2^{1/2}\pi^{1/4}(1-\exp(-\lambda^2 \xi_0^2)^{1/2}}
\biggl\{\exp(\frac{-\xi_0^2 (k+\lambda)^2}{2})
-\exp(\frac{-\xi_0^2 (k-\lambda)^2}{2})\biggr\},
\end{eqnarray}
where $\Delta\xi_1$ is the delay length of the first delay line DL$_1$.
The correlation function is,
\begin{equation}
C=\int_{\infty}^{\infty}dk\beta^*(k)\gamma(k)=
\frac{2^{1/2}}{\{1-\exp(-\lambda^2/\xi_0^2)\}^{1/2}}
\exp\biggl(-\frac{\Delta\xi_1^2+\xi_0^4\lambda^2}{4\xi_0^2}\biggr)
\sin\biggl(\frac{\Delta\xi_1\lambda}{2}\biggr).
\end{equation}
Then, 
\begin{equation}
P_{UU}=I_a+I_c=\frac{1}{4}(1+|C|^2)\sin^2(k_0\Delta\xi_2)
\end{equation}

Figure shows the case of $\lambda\xi_0=1$. The probability has
bumps on both side of $\Delta\xi=0$, where the correlation $C$
is not zero.

Equation~(\ref{small}) shows that $P_{UU}$ is $\sin^2(k_0\Delta\xi_2)$
multiplied by a positive constant regardless of the input photon shapes.
Therefore, for the operation of Sec.~\ref{inputshift} and in this section
we observe always full-swing half-period oscillation. The situation is 
the same even when the input pulse-shape and relative timing  change
at every event, and the observation is the sum of all events.

\bibliographystyle{prsty}

\begin{thebibliography}{99}
\bibitem{simu}
The wave we used in this calculation is as follows.
The input photon has 
Gaussian envelope with random phase variation $\eta(\xi)$. 
\begin{equation}
E(\xi)=\exp\biggl\{-\frac{\xi^2}{2\xi_0^2}+i\eta(\xi)+ik_0\xi\biggr\},
\end{equation}
where
\begin{equation}
\eta(\xi)=\sum_{j=1}^5 3\cos(10 y_j\xi+ 2\pi z_j),
\end{equation}
where $y_j$ and $z_j$ are sequence of random numbers 
between 0 and 1.
\end{thebibliography}

\end{document}